# Jet-induced cratering of a granular surface with application to lunar spaceports

Philip T. Metzger[1], Christopher D. Immer[2], Carly M. Donahue[3], Bruce M. Vu[4], Robert C. Latta III[5], Matthew Deyo-Svendsen[6]

[1]KSC Applied Physics Lab, NASA, Kennedy Space Center, Florida 32899, Philip.T.Metzger@nasa.gov
[2]ASRC Aerospace, Kennedy Space Center, Florida 32899
[3]Department of Physics, Astronomy and Geology, Berry College, 2277 Martha Berry Hwy. NW, Mount Berry, GA 30149
[4]Structures and Analysis, Engineering Development, NASA, Kennedy Space Center, Florida 32899
[5] Department of Aerospace Engineering, Embry-Riddle Aeronautical University, 600 S. Clyde Morris Boulevard, Daytona Beach, FL 32114
[6] Department of Mathematics and Computer Science, Stetson University, 421 N. Woodland Blvd. DeLand, Florida 32723

### *Abstract*

The erosion of lunar soil by rocket exhaust plumes is investigated experimentally. This has identified the diffusion-driven flow in the bulk of the sand as an important but previously unrecognized mechanism for erosion dynamics. It has also shown that slow regime cratering is governed by the recirculation of sand in the widening geometry of the crater. Scaling relationships and erosion mechanisms have been characterized in detail for the slow regime. The diffusion-driven flow occurs in both slow and fast regime cratering. Because diffusion-driven flow had been omitted from the lunar erosion theory and from the pressure cratering theory of the Apollo and Viking era, those theories cannot be entirely correct.

### *Introduction*

During the Apollo and Viking programs there was considerable research into the blast effects of launching and landing on planetary regoliths. That work ensured the success of those missions but also demonstrated that soil erosion or cratering will be a significant challenge for other mission scenarios. For example, the high-velocity spray of eroded soil will pose a serious challenge when we attempt to land multiple spacecraft within short distances of one another on the Moon. We have relevant experience because the Apollo 12 Lunar Module landed 155 meters away from the deactivated Surveyor 3 spacecraft. Portions of the Surveyor were returned by the Apollo astronauts to Earth for analysis. It was found that the surfaces had been sandblasted and pitted and that its openings had been injected with grit from the high-speed spray [Cour-Palais 1972]. This treatment is not acceptable for functional spacecraft.

A program has begun to develop plume/soil mitigation techniques, quantify their effectiveness, and provide the environmental design requirements for hardware near the launch and landing site. It is extremely expensive to perform the experiments using realistic lunar soil and a hypersonic engine plume while maintaining vacuum in the chamber, so we have adopted a strategy that relies heavily on numerical simulations of the physics.

These simulations will precede the more expensive, high fidelity tests that are expected to occur later in the program. Early testing is focused on understanding the physics of plume/soil interactions so that the numerical simulations may be coded properly. There has never been an adequate description of the physical processes or scalings that occur inside a jet-induced erosion event. Also, the prior lunar erosion theories based Shield's parameter were incorrect in that they extrapolated terrestrial experience over many orders of magnitude for flow conditions where different aspects of the physics dominate. Therefore, the space program needs to better understand the basic physics of cratering and erosion processes before it can confidently develop the technology to control them.

The theory developed during the Apollo and Viking programs predicted three different cratering mechanisms, not all of which are applicable to the Moon. The first mechanism is *viscous erosion*, which occurs along the very top layer of sand grains as the dynamic pressure of the gas torques them up and over their neighbors and then pushes them away. This theory as applied to lunar landings was developed primarily by Roberts [1963] and studied experimentally in relation to the Moon by Land and Clark [1965], Hutton [1968], and Clark [1970]. The second mechanism may be called *bearing capacity failure*, and it occurs when the stagnation pressure under the jet exceeds the bearing capacity of the soil and mechanically pushes it downward to form a rather narrow cup. It was studied by Alexander, *et al* [1966] and has not occurred during lunar landings due to the high relative density and shear strength of the lunar soil. It is expected to be a very serious problem for martian landings and launches. The third mechanism is *diffused gas eruption* – largely an auxiliary effect rather than a primary cratering mechanism. It occurs when the dynamic pressure of the jet drives gas into the pore spaces of the soil only to erupt at another location or time, carrying soil with it. This was studied by Scott and Ko [1968]. We shall explain below that two of these mechanisms (viscous erosion and bearing capacity failure) have been partially misinterpreted because the drag of diffusing gas through the soil is a distributed body force that can cause the soil to shear in the bulk. This modifies our understanding of both viscous erosion and bearing capacity failure. This *diffusion-driven flow* may also be viewed as a fourth cratering mechanism in its own right, and the predominant one in many situations.

Erosion and cratering from vertically impinging, circular jets is important to more than just space exploration, including several aerospace, hydraulics and mining applications as well as geology. Several researchers have noted that the crater depth and/or width grow as the logarithm of time over several decades before reaching an asymptotic size and have studied the asymptotic crater profiles and scaling (see the review in Rajaratnam and Beltaos [1977] and Rajaratnam [1982]).

## *Experimental Methods*

This paper describes a set of experiments that use subsonic jets of compressed gas aimed vertically at the surface of dry, cohesionless sand. Two distinctly different cratering regimes were studied. For the higher velocity jets, the cratering process was more energetic and excavated a deep, narrow cavity in the soil with turbulent aggregative fluidization of the sand in the upper part of the cavity. This deep cavity then collapsed into a shallow surface crater when the jet was extinguished. In the lower velocity tests, the cratering process was very smooth, creating a broad, shallow crater by viscous erosion along its inner surface.

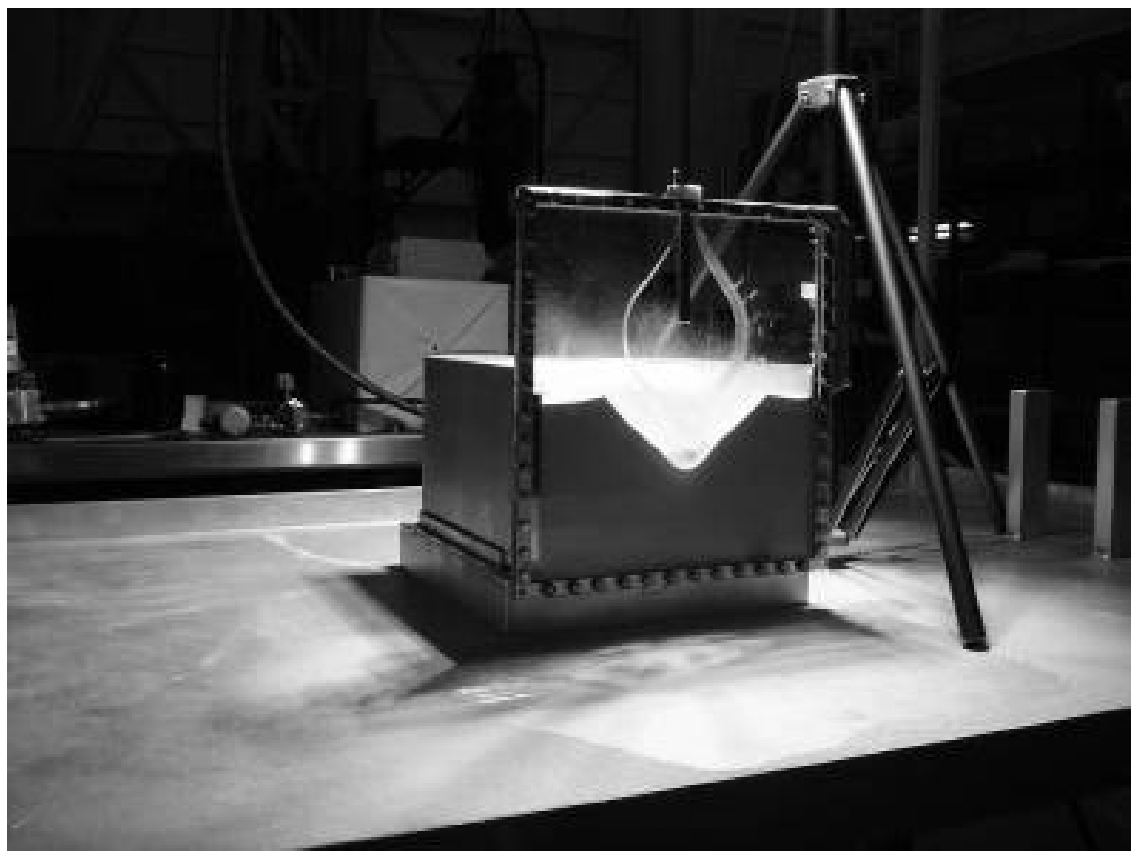

**Figure 1:** Cratering apparatus showing crater formation along the transparent front wall of the box. The cutout in the clear sheet allow the gas to flow freely while keeping most of the sand inside the box

Two different experiment methods were used. In the first method, used only in the high velocity regime, different colors of sand were laid down in thin, horizontal layers. The sand was then cratered by the jet and the jet extinguished. The shallow, collapsed crater was filled in with black sand (to provide a level surface without obscuring the shape of the crater), and the sand was impregnated by optically clear epoxy and cured in an oven. The block of epoxy-stabilized sand was cut in half to reveal how the flat colored layers of sand had been deformed beneath the surface during the cratering process.

The second experiment method was developed by Robert Haehnel [2005] of the U.S. Army Cold Regions Research and Engineering Laboratory. It was used in both the high and low velocity regimes. It uses a plexiglass sheet (beveled outwardly) as the front wall of the sand box, aiming the circular jet so that it straddles the sharp edge of the bevel as shown in Figure 1. Neglecting the drag of the gas along the sheet, this splits the experiment into bisected axial symmetry so that the cratering effects may be observed beneath the surface of the sand as they occur. For the high velocity regime we used high speed videography to capture the cratering event. For the low velocity regime we used high resolution videography. We developed a software algorithm to automatically analyze the high resolution video frame-by-frame throughout the duration of the test to extract crater shape and related parameters and to perform surface and volume integrals on the crater shape. This automated analysis was applied only to the low velocity regime because the turbulent fluidization of the high velocity regime is difficult to analyze by automated methods. The high-speed videos from the high velocity regime were analyzed qualitatively and also by tracking individual sand particles and/or tracer particles that had been mixed with the sand. This allowed features of the sand's velocity field to be discerned. Also, numerical simulations were performed to help interpret both the slow and fast regimes.

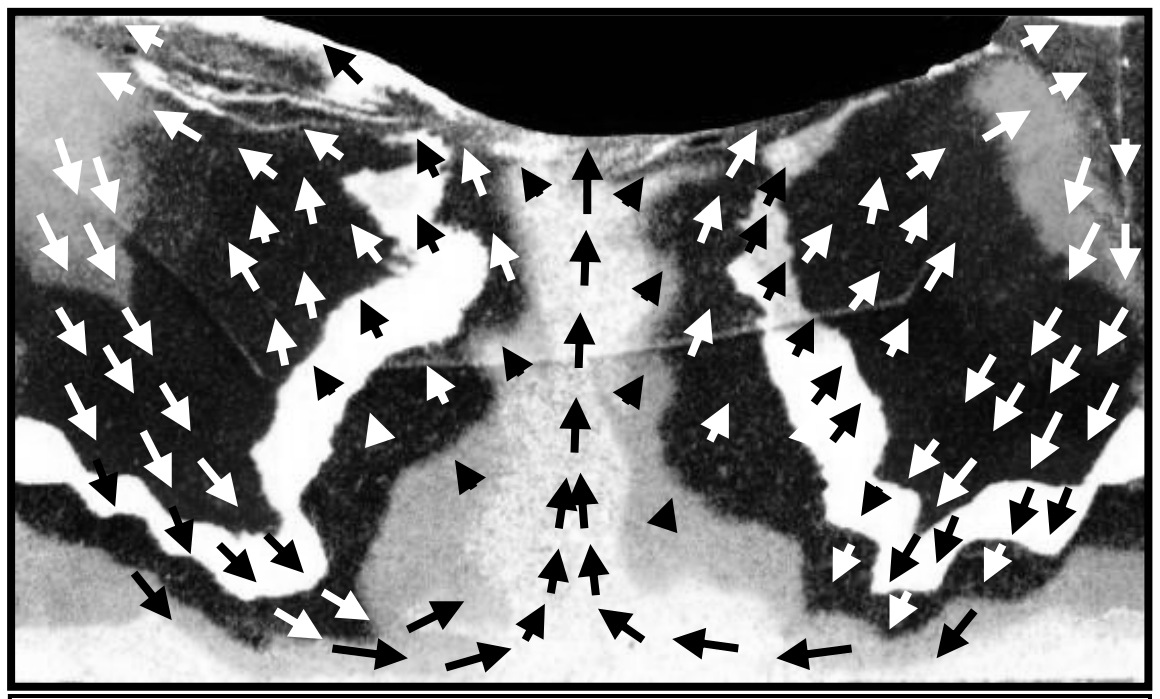

**Figure 2:** Cross-section of epoxy-stabilized sand after cratering event. Layers of colored sand were originally horizontal before the cratering occurred. Arrows were added to show direction of sand flow during the cratering event.

### *Experimental Results:  Fast Regime*

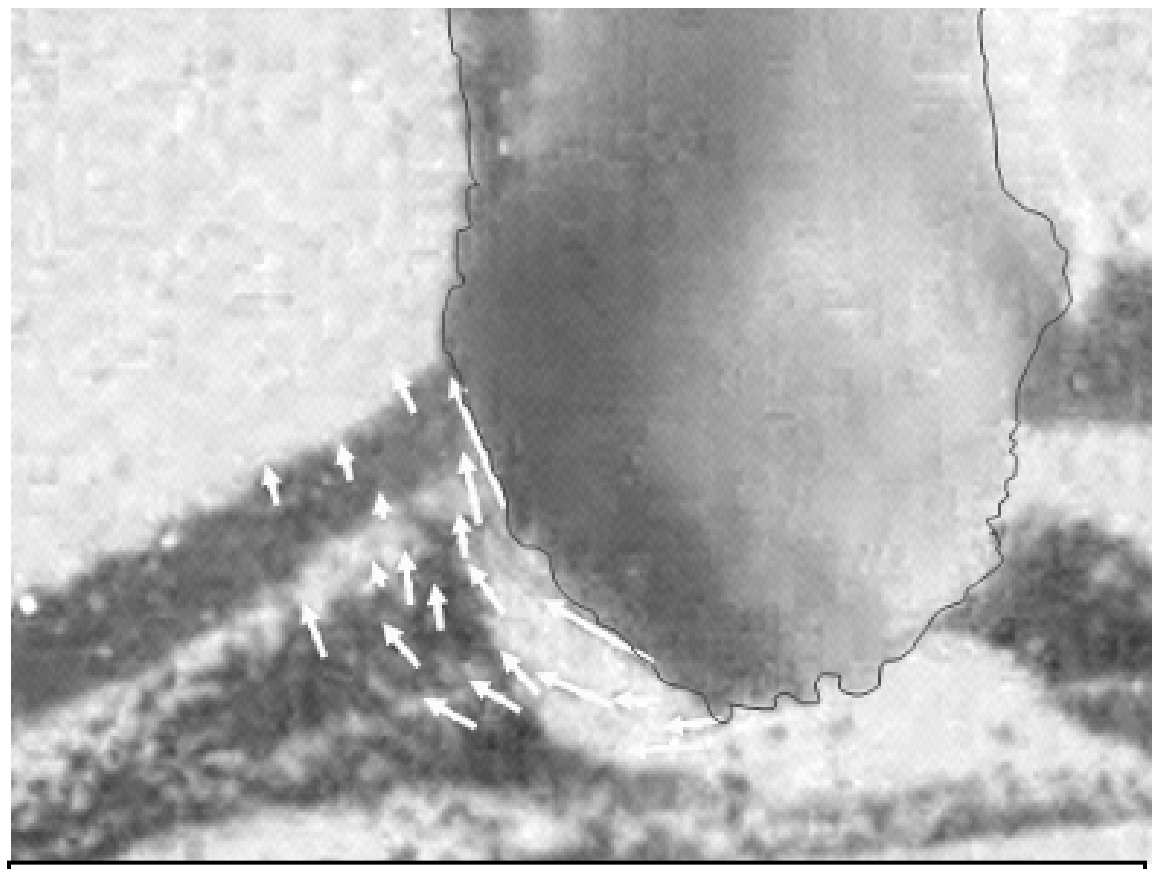

**Figure 3:** Crater (outlined) in white sand with black layers intermixed. White vectors are the velocities of sand particles measured over a small increment of time. Particles travel in the tangential direction around the crater, not in the perpendicular direction as predicted by the theory during the Apollo era.

The Apollo-era theory on the bearing capacity failure mechanism predicted that horizontal, colored layers of sand would be deflected downward beneath the crater due to the pressure of the jet pushing down on the gas-weakened soil. It predicted that the sand would move more or less perpendicularly away from the crater as it grows, creating a sand flow field similar to that which occurs in a cone penetration test. However, when the cratered and epoxy-stabilized sand was split open, we discovered that the deepest layers of sand were actually brought upward to the surface in the center of the crater, rather than being shoved downward. This is shown in Figure 2. Using the second experiment method, it was noted that a narrow cavity of fixed diameter (a little wider than the applied jet) burrows without any apparent resistance vertically downward into the sand until it either stops abruptly by itself or hits the bottom of the container. As the cavity tunnels down, we observed only minimal downward compression of the sand layers directly beneath the cavity's tip. Surprisingly, we did not observe sand being blown upwards and out from the bottom part of the cavity, either. This raised the question where the sand was going so that it made room for the cavity to grow, since it was not being pushed downwards nor was it being removed upwards through the cavity. Individual sand particles were therefore identified and tracked in the video, and it was determined that the sand particles flow in a thick band tangentially around the cavity. This is illustrated in Figure 3. This is contrary to the bearing capacity failure theory of the Apollo-era, which predicted the sand would move perpendicularly away from the cavity's surface. This thick, shearing region of particles moves radially away from the tip of the cavity and then follows the shape of the cavity upward along its sides. The sand eventually exits the bulk by moving into the crater somewhere high up along the cavity's shaft, and it is then swept the rest of the way up and out of the cavity by the gas flow. It is this upward motion in the bulk of the sand tangential to the sides of the cavity that causes the layers to deform upwardly instead of downwardly as had been predicted by the Apollo-era theory.

### *Analysis of the Fast Regime*

To interpret these experimental results, computational fluid dynamics (CFD) simulations were performed using the software package CFD++. Crater geometries were scaled from the videos and used as the stationary boundary conditions for the CFD under the assumption that sand motion is slow relative to the gas attaining its steady state flow above the sand. The boundaries were also treated as impermeable with a no-slip condition, al-though it is known that gases do diffuse through the sand and that the gas velocity across a permeable medium does not strictly obey the no-slip condition at the surface. These approximations are quite adequate considering the high flow of the gas above the sand compared to the very restricted flow through the sand in the bulk and near the surface. The pressures in the cavity as predicted by the CFD are shown in Figure 4. In the shallowest two cavities, the jet impinges directly onto the inside surface of the crater so that high pressure gradients exist in the bottom of the crater. In the deepest two cavities, the core of the jet no longer reaches the bottom of the cavity and so a large plug of high pressure develops in the bottom and the large pressure gradient is located higher up along the shaft. These pressures determine the boundary condition for diffusion of gas into the sand.

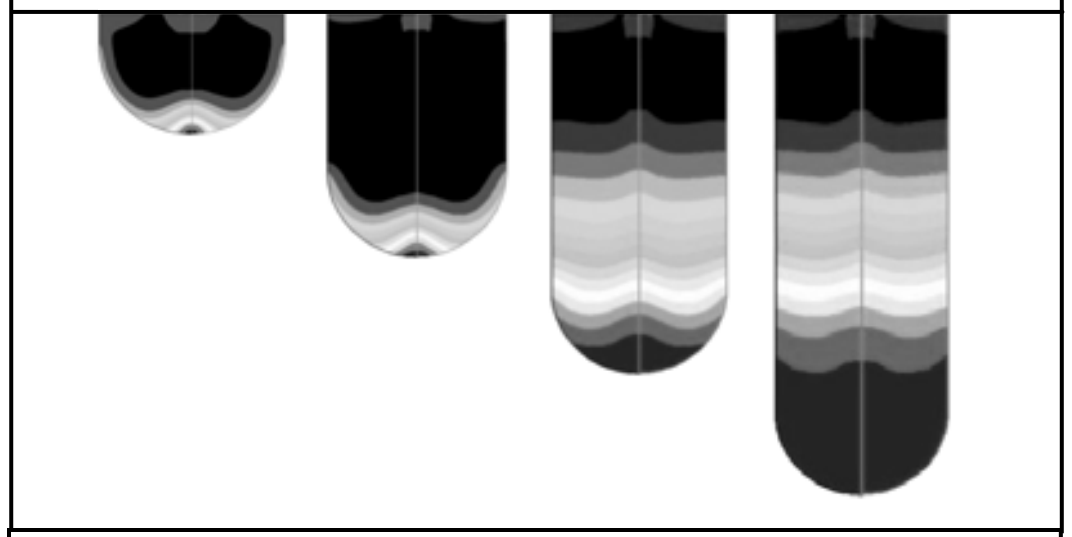

**Figure 4.** CFD simulations during fast-regime crater formation, showing the gradients of gas pressure developed in the crater. The dark regions at the bottom and top of the cavity are the highest and lowest pressures, respectively, and the shades of gray indicating intermediate pressures indicate regions with large pressure gradients. The large gradients are in the crater tip when the crater is small and move up the shaft when it

The diffusion of gas into the pore spaces of the sand is described by Darcy's Law,

$$\vec{v} = \frac{k}{\eta\varepsilon}\vec{\nabla}P$$

where $v$ is the gas velocity field in the sand, $P$ is the pressure field, $\eta$ is the gas viscosity, $\varepsilon$ is the soil porosity, and $k$ is the soil permeability. Combined with the conservation of mass and the assumption of isothermal flow, this produces the equation

$$\nabla^2 P^2 = \frac{2\eta\varepsilon}{k}\frac{\partial P}{\partial t}$$

Defining the characteristic distance $r_0$ to be the nozzle inner radius and the characteristic pressure $P_0$ to be the dynamic pressure of the jet, then the characteristic time for the diffusion is

$$\tau = \frac{2\eta\varepsilon\, r_0^{\,2}}{kP_0}$$

and the characteristic diffusion velocity is $v_0 = (r_0 / \tau)$. The sand used in these tests is called Jetty Park sand and was obtained from the floor of the Atlantic Ocean just offshore from Jetty Park at Cape Canaveral, Florida. It is a fine quartz sand, and it was sieved to retain only the 100-180 micron range. Using the Krumbein and Monk equation [1943], the permeability of the sieved sand is estimated at ~ 4740 Darcies. For typical test parameters this gives a characteristic diffusion velocity of $v_0$ ~ 10 cm/s, and the growth of crater depth is typically on the same order of magnitude. Hence, we cannot assume that gas diffusion is steady state relative to the motion of the sand. We are in a regime where either the diffusive effects of the gas or the undiffused surface pressure (leading to bear-ing capacity failure) might be more dominant, and a mixture of the effects may occur. That is, if the gases diffuse much faster than the sand moves, then there would be negligible mechanical force on the inside surface of the crater and so all motion would be caused by diffusive drag through the bulk of the sand; but if the gases diffuse much slower than the sand moves, then there would be significant mechanical force on the inside surface of the crater and negligible diffusive drag, and so all the motion would be caused by bearing capacity failure. As it is, our approximation of the sand's permeability puts us close to the intermediate region where either process may contribute depending on the exact conditions. In our experiments, we observed that diffusive effects tended to be more dominant with the sand moving tangentially around the crater (rather than perpendicularly as predicted by the bearing capacity theory), although in many cases a mixture of effects were observed with the sand moving neither perfectly tangentially nor perfectly perpendicularly to the crater's surface.

To gain a qualitative understanding of the case where the diffusive effects are dominant, we solve the pressures for the steady state case, $\nabla^2 P^2 = 0$. This was solved in FORTRAN with a finite difference algorithm using the boundary pressures from the CFD and iterating until convergence. The velocity field of the diffusing gas is then obtained by Darcy's Law and the acceleration of the gas is determined by its material derivative. This acceleration is caused by the pressure gradient of the surrounding gas and the flow resistence in the sand, gravity being negligible for the gas. The drag force on the sand is negative the drag force on the gas, and with gravity (which is not negligible for sand), the total body force on the sand is found to be,

$$\vec{f} = \rho_{sand}\ \vec{g} - \rho_{gas}\left(\frac{k}{\eta\varepsilon}\right)^2 \left(\vec{\nabla}P \cdot \vec{\nabla}\right)\vec{\nabla}P + \vec{\nabla}P$$

where $\rho_{sand}$ and $\rho_{gas}$ are the material densities and $g$ is the gravitational acceleration. The middle term, which is the inertial force of the gas, may be neglected as it is many orders of magnitude weaker than the pressure gradient term. The gas pressures and the drag forces in the sand, as predicted by the CFD, are shown for the various crater shapes in Figures 5 and 6, respectively. When the crater is still small and growing, the forces are shown to be tangential around the tip of the crater, perpendicular to the pressure gradient. Therefore, the sand shears in bulk around the tip of the crater (as observed in Fig. 3) to a location that is higher up the sides of the crater. When the sand gets to this higher location the forces push it radially inward, into the crater so that it gets carried away by the gas. In Figure 3 and in other tests we noted that layers of colored sand were deformed upward instead of downward as the crater penetrated through them. This is because the crater grows not by downward mechanical penetration by but the upward fluid-driven shearing. Also, the depth of excavation is limited by the length of the jet's potential core. When the jet's core touches the bottom of the crater, the large radial pressure gradients in the sand enable tangential shearing as illustrated in Figure 5; but when the jet's core no longer reaches to the bottom of the crater, the tangential component of the pressure gradients are large only farther up along the sides of the crater and so tangential shearing beneath the tip of the crater ceases, which arrests crater growth. This explains the observations in the experiments where the craters excavate quickly down to some level and then suddenly their round tip became flattened and excavation ceased. (We also note that if the jet excavated down to the solid bottom of the sandbox, or if it reached a layer of sand that was packed more densely than the sand above it, then the excavated crater might develop a toroidal cavity of rotating gas around the bottom of the crater rather than continuing to excavate downward.)

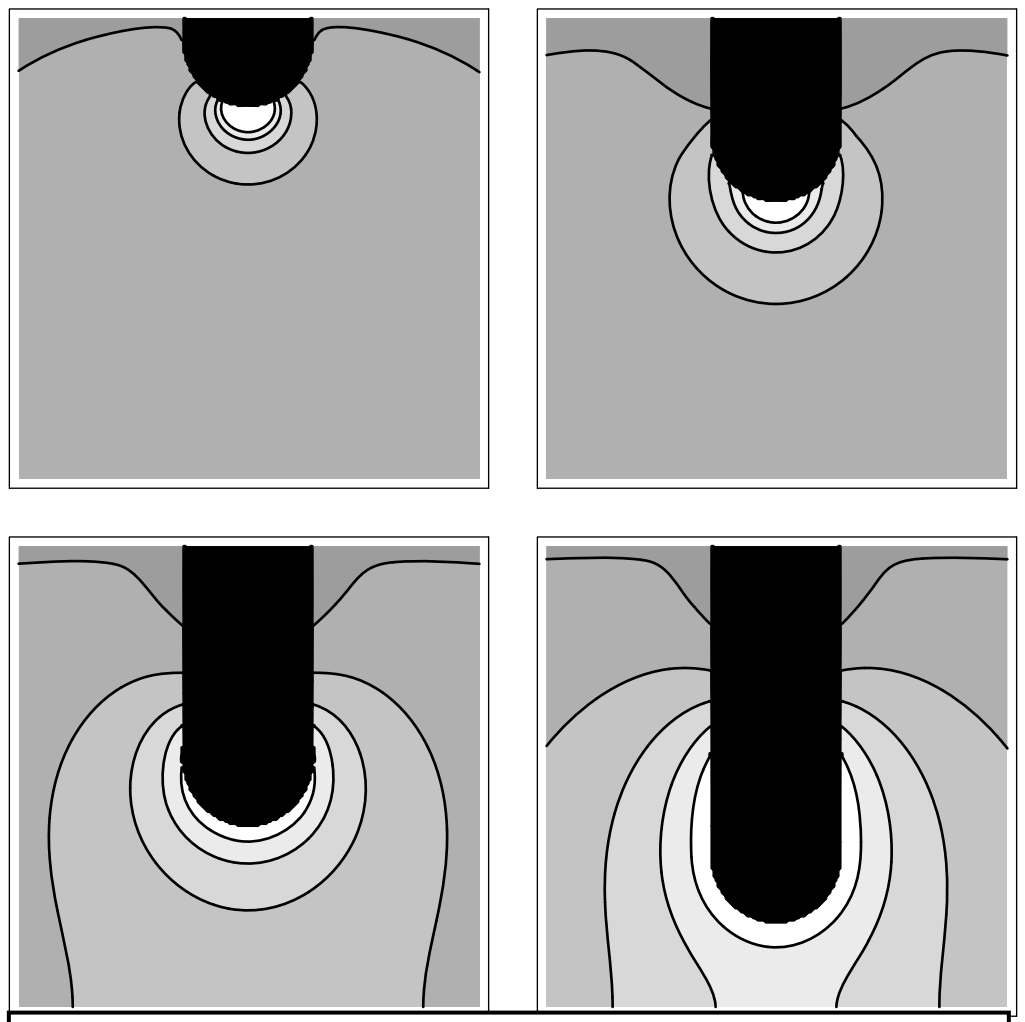

**Figure 5.** Lines of constant pressure within the sand due to diffusing gases, shown in a sequence as the crater grows. The crater is shown in black. The sequence is from top left to top right to bottom left to bottom right.

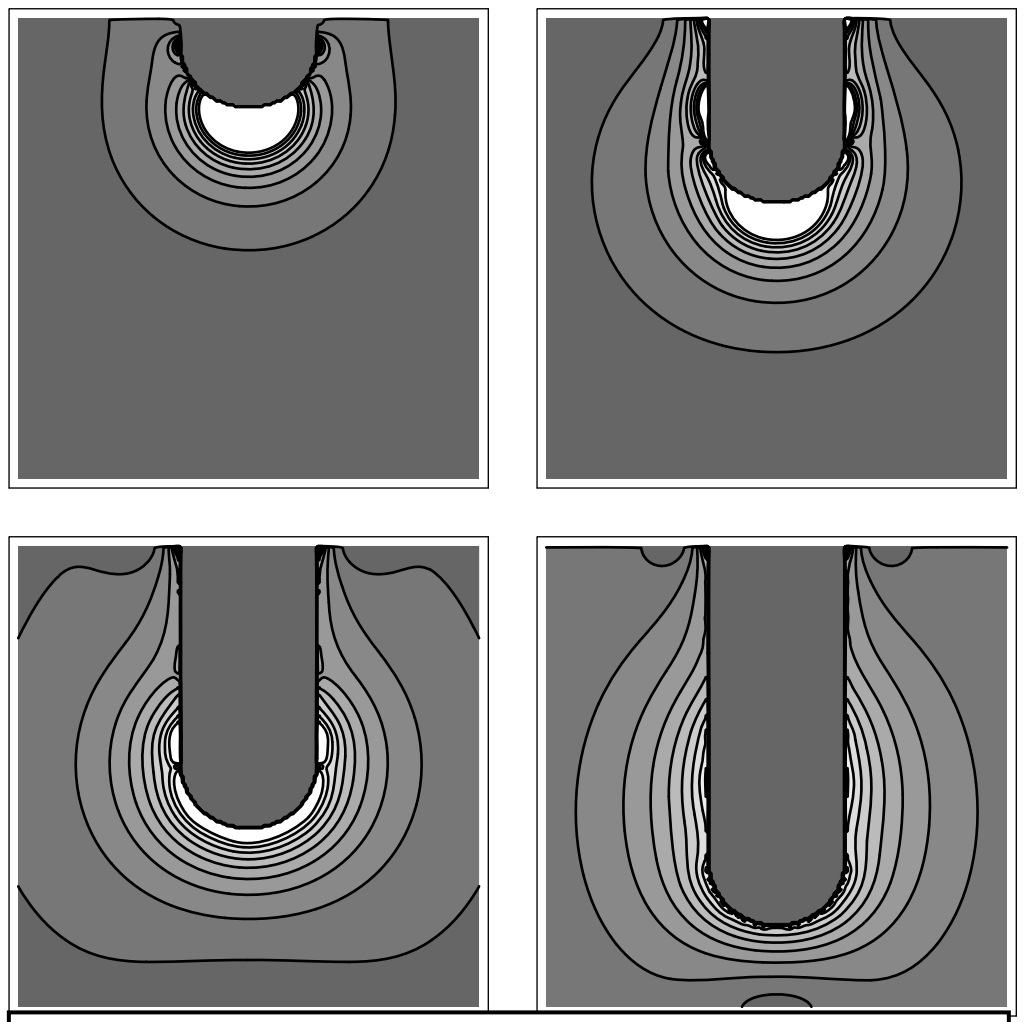

**Figure 6.** Magnitude of the body forces in the sand induced by drag of the diffusing gas, shown in a sequence as the crater grows. Lighter shade indicates stronger force. The direction of the forces is perpendicular to the lines of constant pressure shown in Figure 5. The sequence is from top left to top right to bottom left to bottom right.

In the test represented in Figure 3, the view into the crater during the test was obscured by the exiting sand, and after the jet was shut down the sand collapsed leaving just a shallow surface crater. Therefore, there was no evidence visible at the top surface of the sand that a narrow cavity had been excavated by this shearing process. Examining the colored layers of sand beneath the surface, we could see that they had been dragged upward along the sides of such a cavity prior to collapsing, leaving the form shown in the figure. During the Apollo and Viking era, experiments were performed in this high-velocity, deep-cratering regime, but there was no knowledge that such deep excavation was occurring beneath the soil during those experiments. All that was known was the relatively shallow crater left over at the end of the test. We now realize that in these types of tests where the sand is sufficiently permeable, most of the physics occurs deep beneath the soil by this diffusion-driven shearing, not in a relatively shallow surface crater by bearing capacity failure.

### *Experimental Results: Slow Regime*

Slow regime tests were performed at four different gas exit velocities (37, 40, 50, and 56 m/s), with five different gases for each velocity (helium, neon, nitrogen, argon, and carbon dioxide), at five different jet exit plane heights above the sand (6.35 to 12.70 cm), and with five different exit plane diameters (0.4572 to 1.02108 cm). Tests have been performed with only two different sand grain diameters, but more are proceeding and will be reported later. These results are for the Jetty Park sand sieved to retain 100-180 microns.

In the tests for the slow regime, the main discov-ery is that the rate of crater formation is controlled by the balance of two competing effects: the rate of soil ejection from the crater, and the rate of soil recirculation back into the crater. We found to our surprise that, although the rate of crater growth continually slows during the crater formation process, the rate of soil ejection is essentially constant the entire time. It is not the rate of ejection, but the increasing rate of recirculation, that governs the crater's growth. This is because recirculation is a function of crater shape and size. As the crater grows, increasing amounts of sand fall back into the crater and cannot ultimately escape, and so the net rate of growth continuously slows and eventually stops.

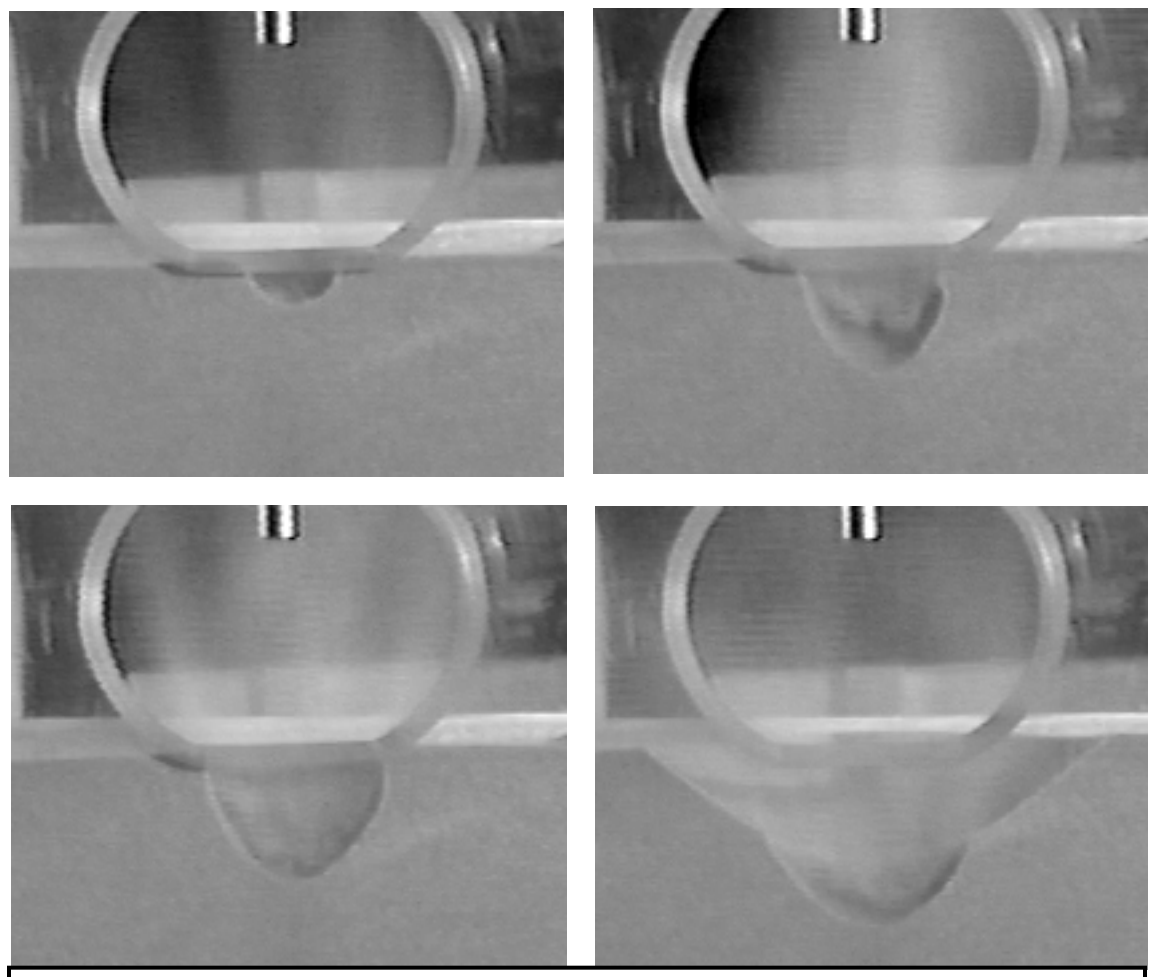

**Figure 7.** Sequence of slow-regime crater growth. (Top left) erosion/fluidization forms a cup. (Top right) Dual lobes of airborne grains forms inside the cup. (Bottom left) Crater widens until traction along inner surfaces is insufficient to maintain steep sides. (Bottom right) Sides collapse to the angle of repose, producing a recirculating surface layer of rolling grains that return to the inner crater.

As first observed by Haehnel [2005], the forming crater evolves into a complex structure illustrated in Figure 7, consisting of a nearly paraboloid inner crater that provides the erosion surface, a conical outer crater at the angle of repose (which is covered by a rolling layer of grains flowing back to the inner crater), a mass of airborne grains re-circulating inside the crater, and dunes of deposited sand around the outer crater's lip.

Figures 8 and 9 show the width and depth of the inner and overall craters versus time for a typical test. The inner crater attains a maximum width and depth and then all subsequent growth of the overall crater is obtained only by widening the outer, conical crater. This is because the inner crater is controlled by gas traction on its inner lip. The upward gas traction sustains the slope of the inner lip at an angle greater than the angle of repose.

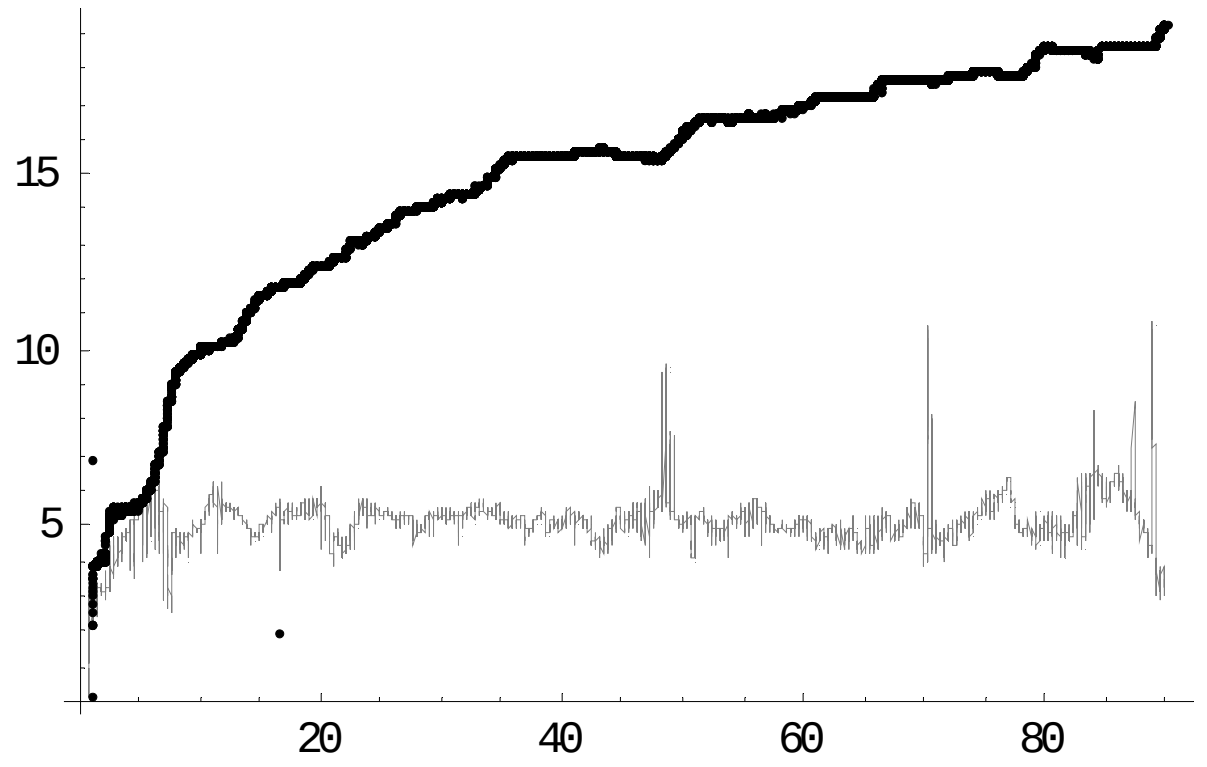

**Figure 8.** Crater Widths (cm) versus time (sec). Outer crater (black); Inner crater (gray). Spikes are due to the video analysis algorithm.

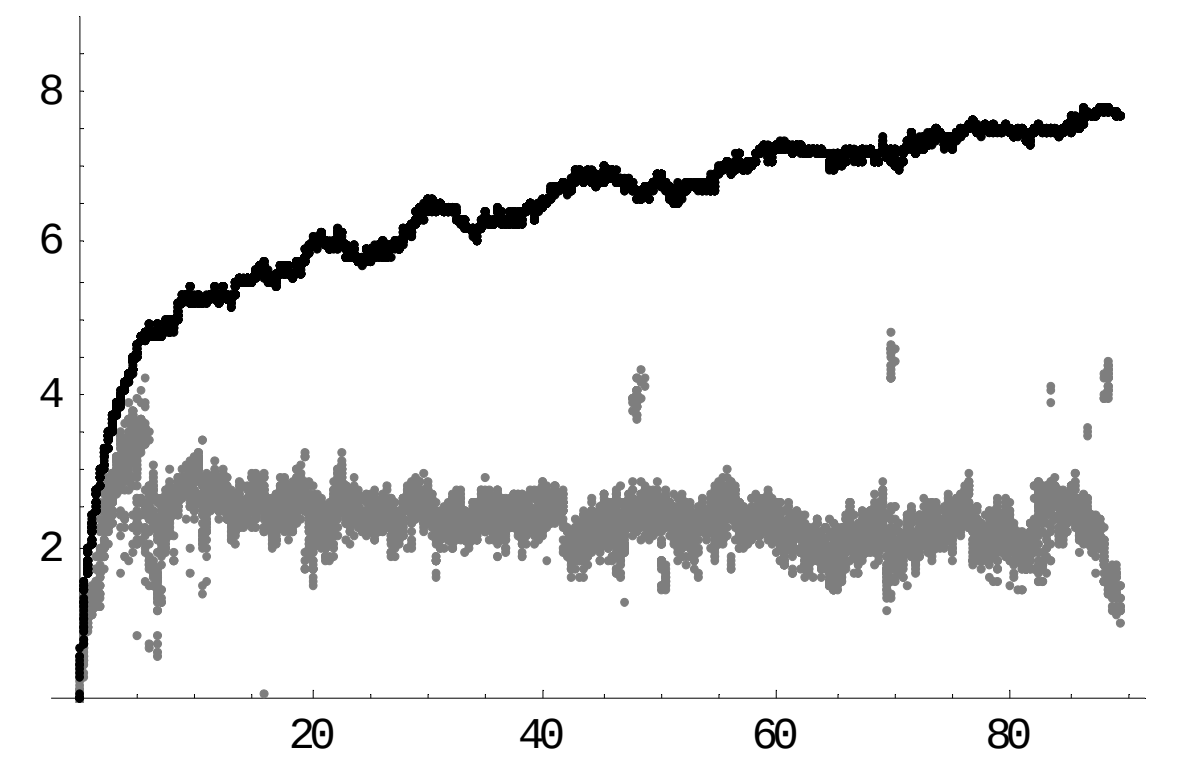

**Figure 9.** Crater Depths (cm) versus time (sec). Overall Crater (black); Inner Crater (gray). Spikes are due to the video analysis algorithm.

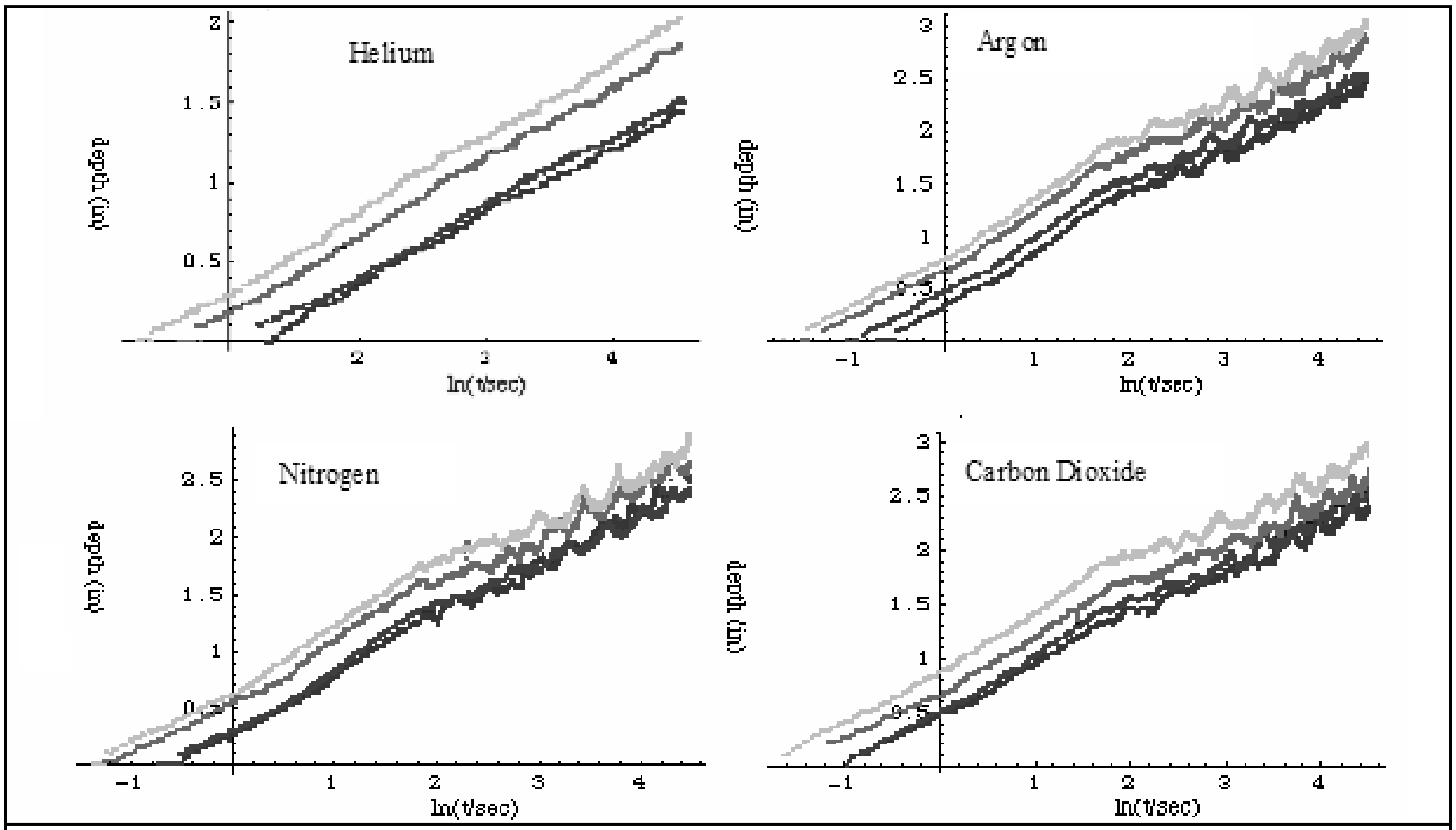

**Figure 10.** Crater depth versus natural log of time in seconds. Each gas is used at exit velocities of 56 m/s (top line, lightest gray), 50 m/s, 40 m/s, and 36 m/s (bottom line, black)

If the crater gets too wide, then the gas exiting the inner crater is spread over too great a circumference and the traction is insufficient to maintain the slope. The lip of the inner crater then collapses until it returns to a sufficiently small circumference to maintain the slope.

As reported by others, the crater depths do grow approximately as the logarithm of time for several decades, as shown by the nearly straight lines in Figure 10. For jets with higher dynamic pressure we observe a knee or an abrupt change of slope in the plots, which corresponds to the moment when the outer crater first begins to form. The plots also begin oscillating after that knee, and the oscillations correlate to periodic avalanching of the outer crater. As the outer crater avalanches, the sand flowing into the inner crater may temporarily impede the excavation, although the exact mechanism of this correlation has not been investigated. The correlation between outer crater avalanches and suspended crater growth is demonstrated in Figure 11.

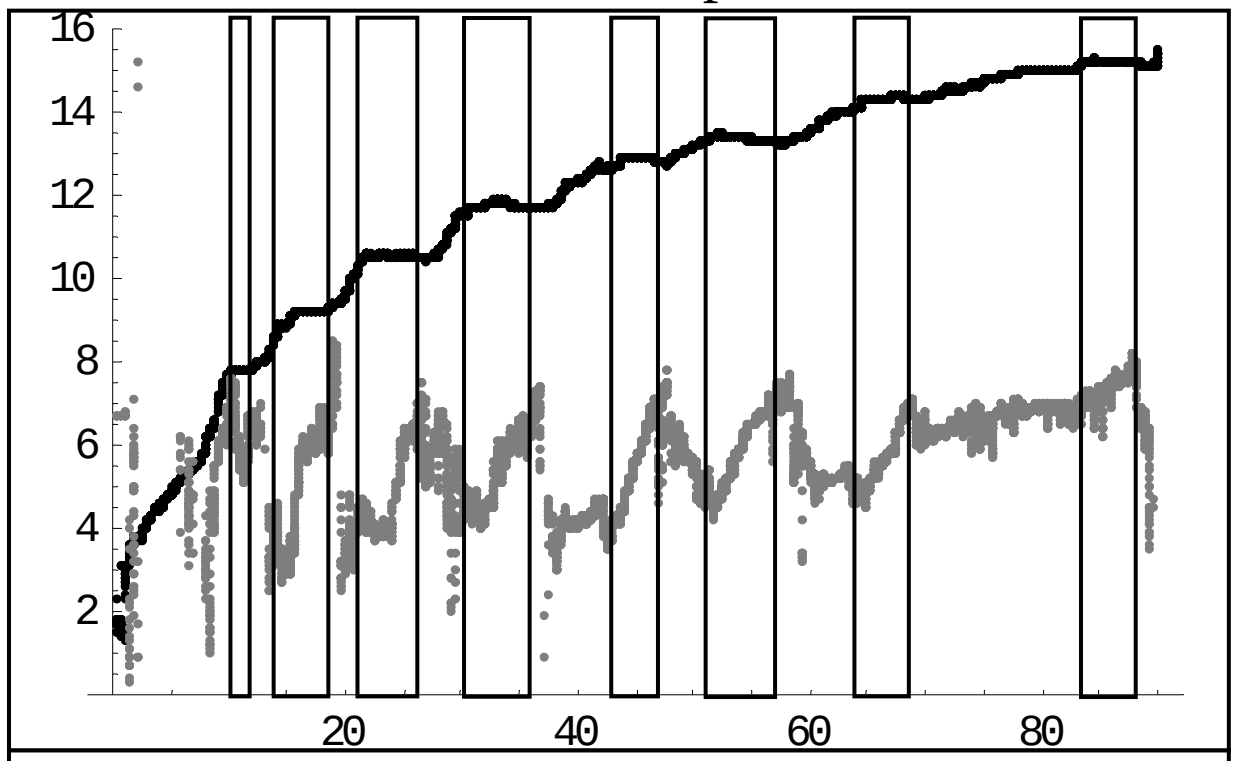

**Figure 11.** Black: crater depth (cm) versus time. Gray: angle of outer crater slope (not to scale) versus time. Vertical bars added as a guide to the eye to show that oscillations in the outer crater angle correlate to oscillations in the crater depth. This is for argon at 37 m/s, nozzle at 7.62 cm height and 1.02108 cm diameter.

The plot of the crater's volume versus time is shown in Figure 12. The plot may be fitted to a power law, $V = \beta\, t^{\alpha}$. For example, with argon at 37 m/s, the nozzle 7.62 cm above the sand and a nozzle diameter of 1.02108 cm, we find that $\beta \approx 9.67$ and $\alpha \approx 0.85$ produces the excellent fit shown in the figure.

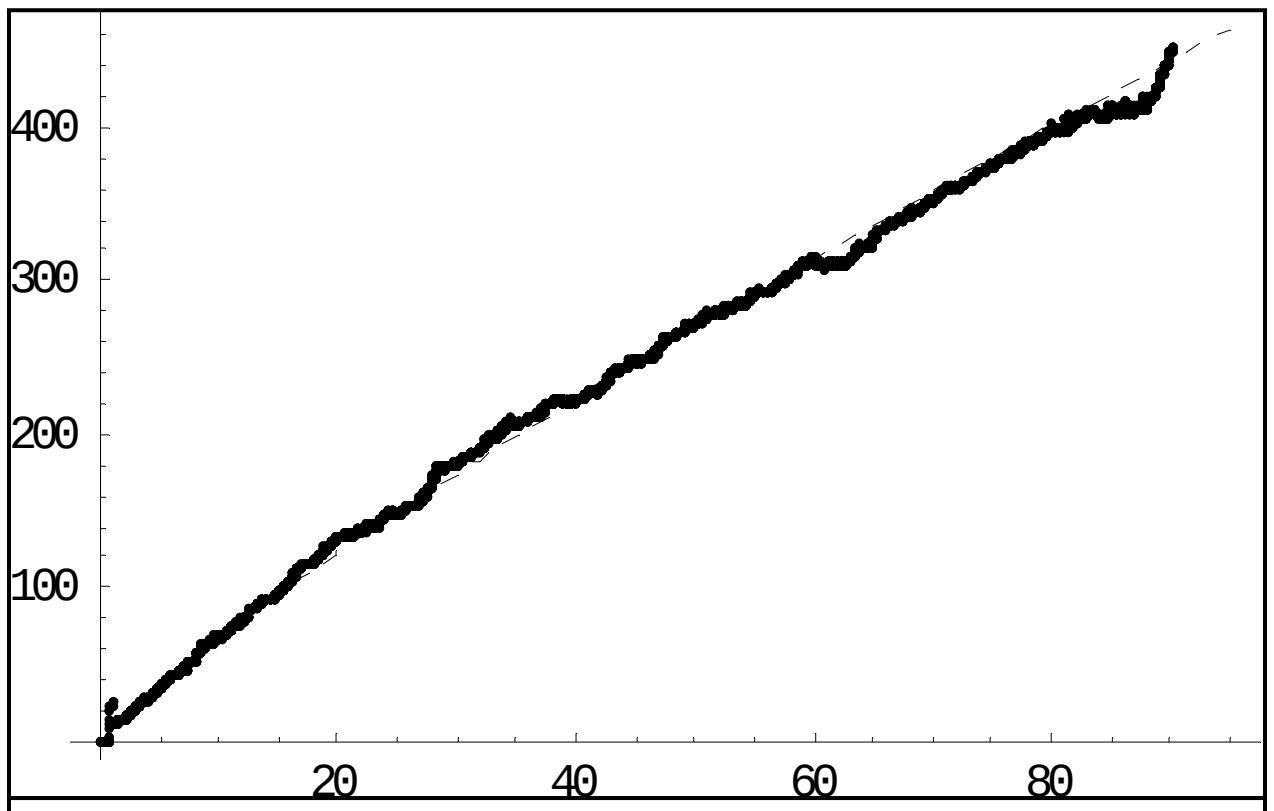

**Figure 12.** Solid line: crater volume (in ml) versus time. Dashed line: power law fit. This is for argon at 37 m/s, nozzle at 7.62 cm height and 1.02108 cm diameter.

As the crater is growing, small "dunes" of the deposited ejecta build up around the lip of the outer crater. These deposits are constantly re-injested by the crater as it widens, and this re-injestion process is oscillatory due to the avalanching nature of the outer crater. Despite the re-injestion of the deposits, their net volume is an essentially monotonically increasing quantity throughout the entire cratering event. The small oscillations are superimposed on this essentially monotonic form as shown in Figure 13 (left). When the deposited volume is subtracted from the crater volume, the difference is the quantity of sand thrown completely out from the sand box. As shown in Figure 13 (right), this volume of lost sand increases linearly with time throughout the entire event, indicating that the loss of sand occurs at a constant rate. This is the only measurement that we have discovered to be linear with time. The implication is that the rate at which sand is thrown into the air is also constant (although we have no easy way to directly measure it). Therefore, the decreasing growth rate of the crater is not the result of a diminishing sand ejection rate – which is a constant – but rather an increasing sand recirculation rate. The more often the same sand must be thrown into the air before it finally falls outside the width of the outer crater, then the slower the net growth rate of the crater becomes.

Prior researchers have studied the size of the asymptotic crater scaled according to the physical parameters such as jet velocity, exit plane height and diameter, etc. The asymptotic size of scour holes is important for certain civil engineering and geological applications. For launching and landing rockets and for certain other aerospace applications the transient dynamics are more important than the asymptotic size. Therefore, we have in-vestigated the scaling of the crater growth rates. This is more important to martian landings than to lunar ones, since this sort of deep cratering (either fast or slow regime) is not expected to occur in the very dense, competent lunar soil. The crater depth is interesting due to the simple, nearly linear growth that one observes with the logarithm of time, and because it has been much studied in the past. We have fitted the depth curves from our experiments to the form $d = a\log(bt+1)$, so that *a* determines the depth scale and *b* provides the time scale. As shown in Figure 14, $a \propto H$ and $b \propto \rho v^2 D / H^{\alpha}$ where *D* is jet exit plane diameter and *H* is its height above the uncratered surface of the sand. The value of $\alpha$ cannot be accurately determined without more data, but it appears to lie between 1 and 5. Surprisingly, the depth scale of the crater is entirely independent of *ρ*, *v*, and *D*. The scaling of *b* is not consistent with the "Erosion Parameter" of Rajaratnam and Beltaos [1977], which was used to scale the crater size in its asymptotic (final) state rather than during its growth. Analysis of *a* and *b* as a function of sand grain diameter, specific gravity, and gravitational acceleration is still on-going. Any possible dependence on gas viscosity would be difficult to investigate here since at room temperature these five gases have similar viscosity. Viscosity is expected to play a greater role in determining whether bearing capacity failure or diffusion driven flow will dominate in the high velocity regime. However, viscosity does play at least some role in the slow velocity regime because diffusion driven shearing does occur beneath the tip of the inner crater under the stagnation pressure of the jet. Since in this slow regime the ejection of sand into the air actually occurs at the lip of the inner crater by viscous erosion, not beneath its tip where diffusion driven shearing occurs, it may be that viscosity plays a greater role in the shape of the inner crater than in the volumetric erosion rate.

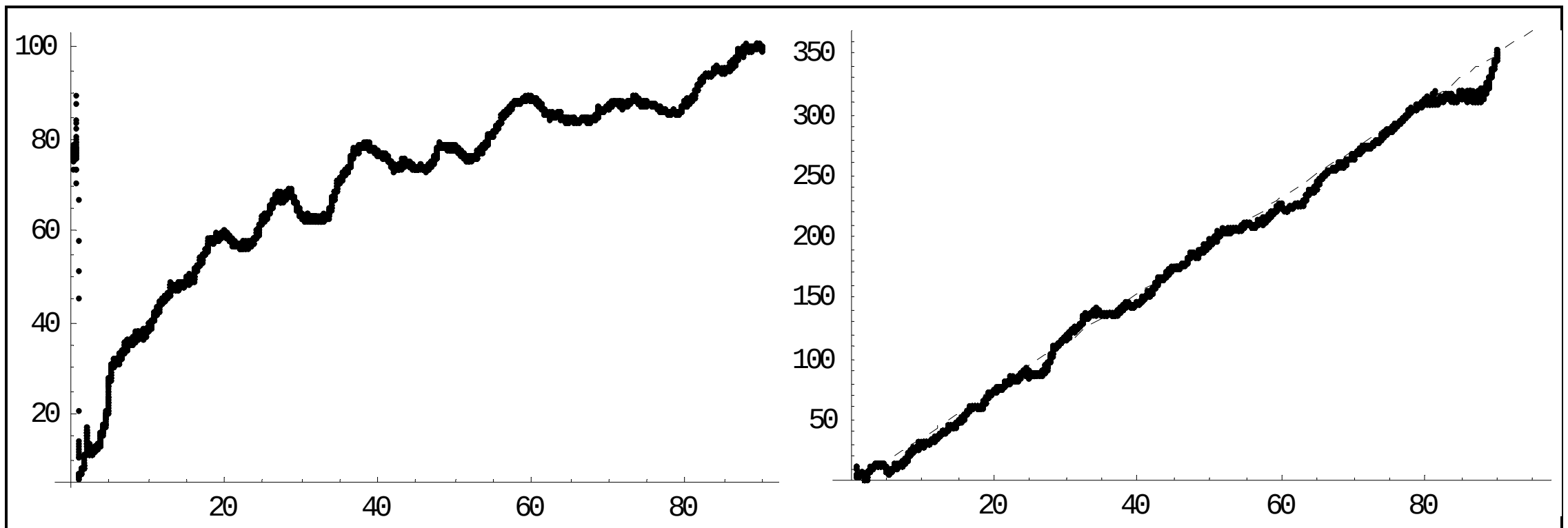

**Figure 13.** Left: Volume of ejecta deposits (ml) versus time (sec). Right: Difference between crater volume and ejecta deposit volume (ml) versus time (sec). The dashed line is a guide to the eye to demonstrate linearity with time. This is for argon at 37 m/s, nozzle at 7.62 cm height and 1.02108 cm diameter.

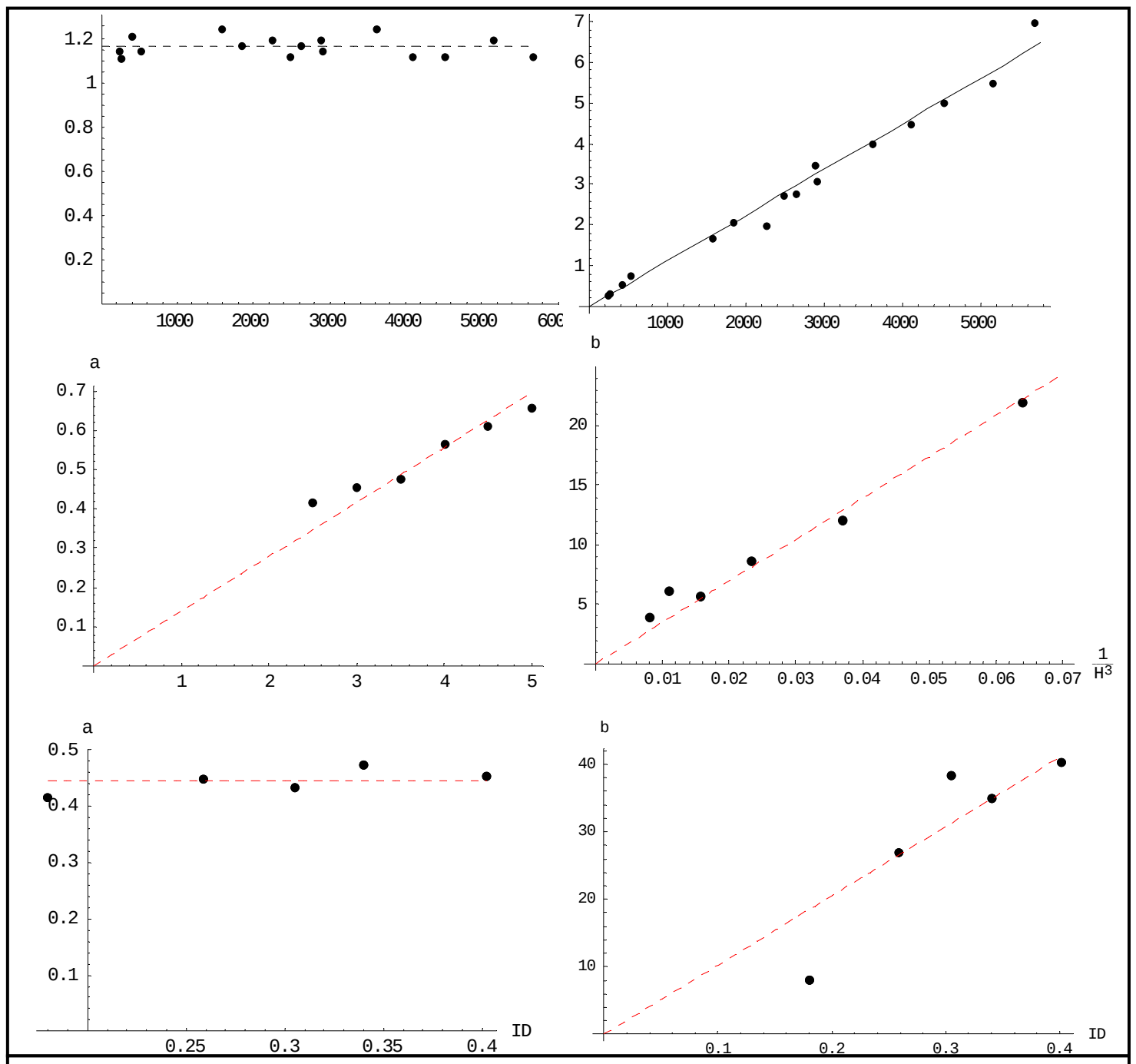

**Figure 14.** Dependence of parameters *a* and *b* on various physical parameters. Top two frames: versus gas density times velocity squared (in Pascals). Middle two frames: versus nozzle height in cm. Bottom two frames: versus nozzle inner diameter in cm.

## *Conclusions*

The following picture of crater formation has emerged. Grains are ejected and then they are either recirculated in the crater or else they escape from the crater. The ejection occurs through the following mechanisms. First (referring to the sequence of pictures in Figure 7), diffusion driven shearing occurs just beneath the tip of the crater where the gases are stagnant, but viscous surface erosion occurs over the rest of the inner cup, rolling the grains uphill until they reach the lip of the inner crater. This was observed in the tests and confirmed by the CFD simulation shown in Figure 15 where maximum pressure gradients are beneath the tip but high velocities are everywhere else. Second, the actual erosion of sand, where it becomes airborne, occurs at the lip at the top of the inner cup. This, too, was observed in the tests and confirmed by the CFD simulation shown in the left side of Figure 15 where the maximum gas velocity occurs at the lip. Third, as the inner cup grows as shown in the sequence of Figure 7, it becomes deeper with steeper sides until the gas traction becomes insufficient to maintain the steep slope and it begins to avalanche, forming the outer crater. This outer crater is completely shielded from the gas flow as shown by the CFD in Figure 15 and so particle shape, size distribution and cohesion determine the slope. We confirmed through measurements that this slope is the usual angle of repose for this sand. Once these outer crater processes begin, there will be two mechanisms that deliver sand to the erosion spot at the lip of the inner crater. The first is shearing and rolling grains in the inner cup to bring them up to the erosion spot. The second is the avalanching in the outer crater to bring the sand down to the erosion spot. After being ejected at this lip, the grains recirculate through the following three mechanisms. First, *aerial recirculation*: some grains are placed into trajectories that drop them directly into the inner crater. Second, *rolling recirculation*: some grains fall onto the conical slopes of the outer crater and are funneled back to the erosion spot where they are ejected again. Third, *deposit recirculation*: some grains fall just outside the outer crater where they form small deposits, but obtain only a temporary respite from the recirculation because the outer crater widens and eventually re-ingests them. The grains that *escape* the crater are the ones that fall so far away that they are never re-ingested. The volumetric growth of the crater is the difference between the volume of sand ejected and the volume of sand recirculated. When the two rates become equal, crater growth stops at the asymptotic size.

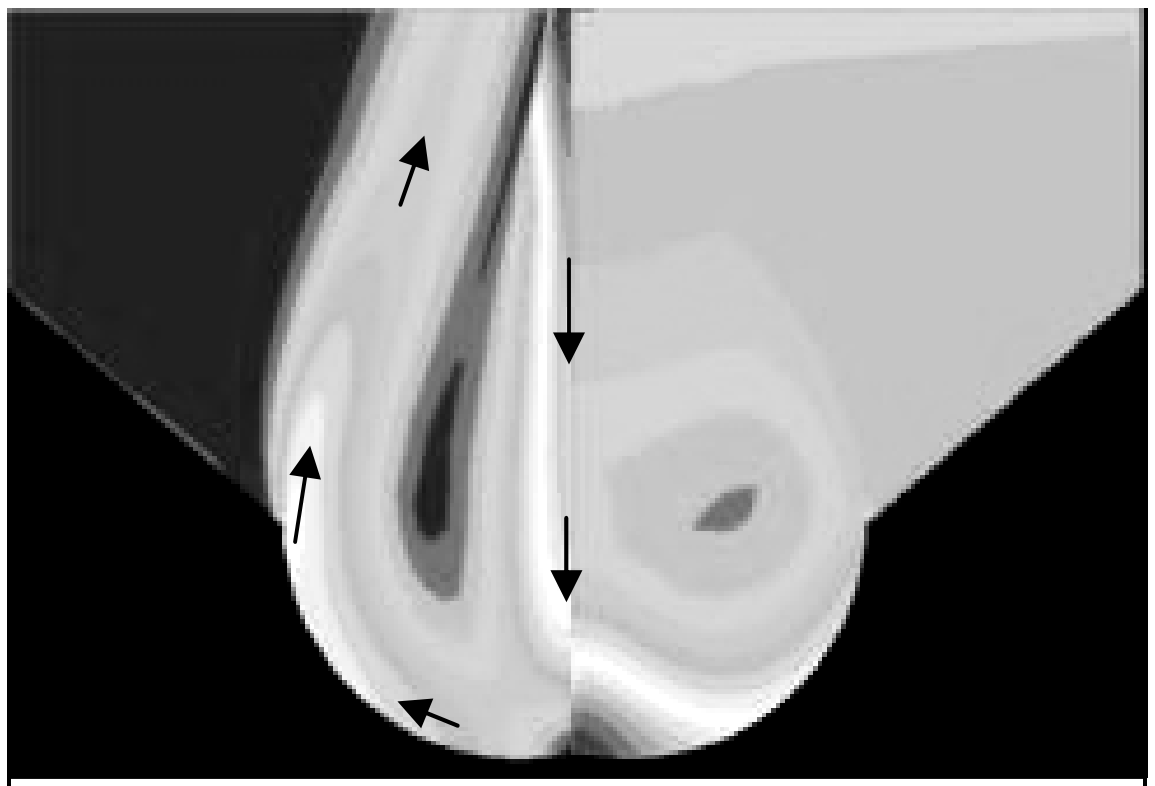

**Figure 15.** CFD of gas flow in slow-regime crater. (Left half) Gas velocity, with lighter shading for faster flow. Arrows added to indicate direction. (Right half) Gas pressure, with high stagnation pressure at the bottom center of the inner crater rapidly reducing to ambient pressure in the radial direction.

Despite the obvious differences, Figure 15 indicates how the crater formation is analogous to lunar soil erosion if the curved surface of the inner crater were flattened into a plane. Then, both the bell-shaped pressure distribution under the jet and the annular erosion ring where the traction is highest would appear remarkably similar to the predictions of Roberts' lunar erosion theory [1963]. This leads to an important observation. Roberts' theory did not include the diffusion driven flow, which delivers sand from the centerline of the jet to the annular region where the sand becomes airborne. In these small-scale experiments, without the diffusion driven flow to move sand away from the stagnation region, the crater would be shaped like a "W" instead of a "U" in the inner crater. Likewise, in the lunar case the mass erosion rate predicted by Roberts' theory cannot be correct if diffusion driven flow is appreciable. At the present it is difficult to predict how significant the diffusion driven flow may be in the lunar environment because we lack an accurate prediction of the permeability in the loosest layer of soil. However, we are currently extending Roberts' theory and hope to have reasonable estimations of permeability and mass flow in the near future. We are also undertaking a series of tests with supersonic jets and developing numerical simulations of different aspects of the erosion process including full-scale integrated numerical simulations of various scopes and fidelities. These experiments and simulations are being used together to develop further insight and to mature the simulation techniques until a truly predictive lunar code has been developed.

### ***Bibliography***

Alexander, J. D., W. M. Roberds, and R. F. Scott (1966). "Soil Erosion by Landing Rockets." Contract NAS9-4825 Final Report. Hayes International Corp., Birmingham, Alabama.

Clark, Leonard V. (1970), "Experimental Investigation of Close Range Rocket Exhaust Impingement on Surfaces in a Vacuum," NASA TN D-5895

Cour-Palais, B. G. (1972), "Part E. Results of Examination of the Returned Surveyor 3 Samples for Particulate Impacts," Analysis of Surveyor 3 material and photographs returned by Apollo 12 (Washington D. C.: NASA), pp 154-67.

Haehnel, Robert (2005) personal communication.

Hutton, Robert E. (1968), "Comparison of Soil Erosion Theory with Scaled LM Jet Erosion Tests," NASA-CR-66704.

Krumbein, W. C., and Monk, G. D., (1943), "Permeability as a function of the size parameters of unconsolidated sand," Transaction of the American Institute of Mining, Metallurgical and Petroleum Engineers, v. 151, p. 153-163.

Land, Norman S. and Leonard V. Clark (1965), Experimental Investigation of Jet Impingement on Surfaces of Fine Particles in a Vacuum Environment, NASA Technical Note D-2633 (Hampton, Va.: Langley Research Center).

Rajaratnam, Nallamuthu (1985), "Erosion by Submerged Circular Jets," Journal of the Hydraulics Division, 108(2), 262-267.

Rajaratnam, Nallamuthu, and Spyridon Beltaos (1977), "Erosion by Impinging Circular Turbulent Jets," Journal of the Hydraulics Division, 103(10), 1191-1205.

Roberts, Leonard (1963), "Visibility and Dust Erosion During the Lunar Landing," A Compilation of Recent Research Related to the Apollo Mission, (Hampton, Va.: Langley Research Center), pp. 155-170.

Scott, Ronald F., and Hon-Yim Ko (1968), "Transient Rocket-Engine Gas Flow in Soil," AIAA Journal 6(2), 258-64.